\documentclass[aps,pre,twocolumn,superscriptaddress]{revtex4}
\usepackage{epsfig}
\usepackage{graphicx}
\usepackage{dcolumn}
\usepackage{bm}
                                                                                                
\begin{document}
\title{Modified Statistical Treatment of Kinetic Energy in the Thomas-Fermi Model}                                    
\author{Jeng-Da Chai\footnote{corresponding email: jdchai@wam.umd.edu}}
                                                                                                                            
\affiliation{Institute for Physical Science and Technology,}
\author{John D. Weeks}
\affiliation{Institute for Physical Science and Technology,}
\affiliation{and Department of Chemistry and Biochemistry, University of Maryland, College Park, MD 20742}
                                                                                                                            
\date{\today}                                                                                                                            
\begin{abstract}
We try to improve the Thomas-Fermi model
for the total energy and electron density of atoms and molecules by
directly modifying
the Euler equation for the electron density, which we argue is less affected by
nonlocal corrections.  Here we consider the
simplest such modification by adding a linear gradient term to the Euler
equation. For atoms, the coefficient of the gradient
term can be chosen so that the correct exponential decay constant far from the
nucleus is obtained. This model then gives a much improved description of the
electron density at smaller distances, yielding in particular a
finite density at the
nucleus that is in good qualitative agreement with exact results. The cusp
condition differs from the exact value by a factor of two. Values for the total
energy of atomic systems, obtained by coupling parameter integration
of the densities
given by the Euler equation, are about as accurate as those given by very best
Thomas-Fermi-Weizs\"{a}cker models, and the density is much more accurate.
Possible connections to orbital-free methods for the kinetic-energy functional
in density functional theory are discussed.
\end{abstract}
\newpage
                                                                                                                    
                                                                                                                    
\maketitle
\section{Introduction}
                                                                                                                    
Density functional theory (DFT) has proved to be a powerful way to determine
equilibrium and dynamic properties of atoms, molecules, and extended systems
\cite{Parr, Dreizler, Car, Kresse, Chai}. The modern conceptual framework
was established by Hohenberg and Kohn (HK) \cite{Hohenberg}, who proved that
the exact ground state energy of a system of $N$ electrons is a functional $%
E[\rho ]$ of the ground state electron density $\rho ({\bf r})$ only, where
\begin{equation}
E[\rho ]=\int \rho ({\bf r})V_{ext}({\bf r})d{\bf r}+F[\rho ],  \label{eq:2}
\end{equation}
and $F[\rho ]$ is an unknown but universal functional of the density and is
independent of the external potential $V_{ext}({\bf r})$.
                                                                                                                    
Kohn and Sham (KS) \cite{Kohn, Sham} showed that $F[\rho ]$ can be usefully
partitioned into two dominate (kinetic and potential energy) terms and a
residual correction:
\begin{equation}
F[\rho ]=T_{s}[\rho ]+\frac{1}{2}\int \frac{\rho ({\bf r^{\prime }})\rho (%
{\bf r})}{|{\bf r}-{\bf r^{\prime }}|}d{\bf r^{\prime }}d{\bf r}+E_{xc}[\rho
].  \label{eq:3}
\end{equation}
The first term on the right of eq (\ref{eq:3}) is the KS {\em kinetic
energy density functional} $T_{s}[\rho ].$ This gives the kinetic energy of
the ground state of a reference or model system of $N$ noninteracting
electrons in a self-consistent field chosen so that the ground state density
equals $\rho ({\bf r})$. The second term is the classical (Hartree) {\em %
electron-electron potential energy} (atomic units are used in this paper), and the last term
is the {\em %
exchange-correlation density functional} $E_{xc}[\rho ,]$ which formally
accounts for {\em all} remaining smaller corrections.
                                                                                                                    
If these functionals were known, then one could obtain the density $\rho (%
{\bf r})$ from the variational principle (Euler equation) associated with
eq (\ref{eq:2}):
\begin{equation}
\mu =V_{T_{s}}({\bf r;[}\rho ])+V_{eff}({\bf r;[}\rho ]),  \label{eq:4}
\end{equation}
and then determine the total energy of the inhomogeneous system from the
energy functional $E[\rho ]$. All other physical quantities related to the
ground-state density could also be determined. Here
\begin{equation}
V_{T_{s}}({\bf r;[}\rho ])\equiv \delta T_{s}[\rho ]/\delta \rho ({\bf r})
\label{kineticpotential}
\end{equation}
is the KS {\em kinetic potential} \cite{KingHandy}, formally defined
as a functional
derivative of $T_{s}[\rho ],$ $\mu $ is the chemical potential (the Lagrange
multiplier associated with the normalization condition $\int \rho ({\bf r})d%
{\bf r}=N)$, and $V_{eff}({\bf r;[}\rho ])$ is an effective one-body
potential for the KS partitioning, defined by
\begin{equation}
V_{eff}({\bf r;[}\rho ])\equiv V_{ext}({\bf r})+\int \frac{\rho ({\bf %
r^{\prime }})}{|{\bf r}-{\bf r^{\prime }}|}d{\bf r^{\prime }}+\frac{\delta
E_{xc}[\rho ]}{\delta \rho ({\bf r})}.  \label{eq:5}
\end{equation}
                                                                                                                    
Through the efforts of many workers, we now have rather accurate expressions
for the {\em exchange-correlation potential} $V_{xc}({\bf r;[}\rho ])\equiv
\delta E_{xc}[\rho ]/\delta \rho ({\bf r})$ in eq (\ref{eq:5}), and the
main problem in solving the Euler equation (\ref{eq:4}) is in finding an
accurate approximation for the kinetic potential as a functional of the
density \cite{YWang3, Parr}. The most straightforward way
to proceed would be to develop
approximations for $T_{s}[\rho ]$ itself, and obtain $V_{T_{s}}({\bf r;[}
\rho ])$ by functional differentiation, but this has proved to be a very
challenging problem. As discussed below, simple local density approximations
for $T_{s}[\rho ]$ based on the Thomas-Fermi (TF) theory are not
very accurate, and make several qualitatively incorrect predictions.
                                                                                                                    
To bypass such problems, KS suggested a different way to compute the
numerical value of $T_{s}$ exactly, not directly from the density itself,
but by introducing a set of $N$ orbitals satisfying the $N$ coupled KS
equations that describe the model system \cite{Kohn, Sham}. Using these
orbitals one can determine both $T_{s}$ and the total density $\rho
({\bf r}) $.
                                                                                                                    
This accurate treatment of the kinetic energy played a central role in the
development of DFT as a quantitative method \cite{Parr}. The KS method
has yielded very
accurate ground-state densities and energies for many systems using
currently available $E_{xc}[\rho ]$. In many cases even a simple local
density approximation for $E_{xc}[\rho ]$ proves adequate, and we will use
this in the results reported here. However, the numerical cost of
self-consistently determining $N$ orbitals rapidly increases for large $N$.
An accurate treatment of the kinetic energy as well as the potential energy
contributions in terms of the electron density only, as originally
envisioned in the work of HK, would certainly be desirable.
                                                                                                                    
To that end, there has been renewed interest in developing ``orbital-free''
(OF) approximations for the kinetic energy density functional $T_{s}[\rho ]$,
with some notable successes in describing extended systems
\cite{YWang3, L-WWang, YWang, YWang2, Madden, Kaxiras}. The main
advance has been to introduce {\em nonlocal} density functionals that
reproduce known exact results for the linear response of the density of the
model system to small perturbations. Similar ideas led to the development of
nonlocal weighted density functional theories for classical nonuniform
fluids, again with some notable successes \cite{Evans}. However, in
both cases it has
proved difficult to gain a physical understanding of how best to incorporate
essential nonlocal effects directly into the kinetic energy (or classically,
the free energy) functional through various weighting kernels and to
understand what errors in the density will likely arise from given
approximations to these functionals.
                                                                                                                    
We pursue here an intermediate strategy between the ultimately desirable
goal of determining $T_{s}[\rho ]$ as a functional of the density and the KS
method, which avoids direct use of the density and uses orbitals to
calculate numerically both $T_{s}$ and the density. We focus instead on
approximating the kinetic potential $V_{T_{s}}({\bf r;[}\rho ])$ as a
functional of the density. The connection to the density response is more
direct and physically suggestive and we believe that nonlocal effects in $%
V_{T_{s}}({\bf r;[}\rho ])$ could be less problematic than in the full $%
T_{s}[\rho ].$
                                                                                                                    
\section{Nonlocal effects in $T_{s}[\rho ]$ and $V_{T_{s}}({\bf r;[}\rho ])$}
                                                                                                                    
$T_{s}[\rho ]$ can be formally related to $V_{T_{s}}({\bf r;[}\rho ])$ by a
(functional) integration over the density changes in all space, as the
density is changed from a uniform to the final nonuniform value
\cite{Chen, Weeks}. Because of
this spatial integration, $T_{s}[\rho ]$ is a more nonlocal functional of
the density than is $V_{T_{s}}({\bf r;[}\rho ])$. Other detailed theoretical
arguments suggesting that $V_{T_{s}}$ is more local have recently
appeared \cite{Holas}.
A second consequence of this integration is that the numerical value of $%
T_{s}[\rho ]$ depends mainly on smoothed or spatially averaged properties of
$\rho ({\bf r}).$ Thus the inverse problem of accurately determining $\rho (%
{\bf r})$ from eq (\ref{eq:4}) at all values of ${\bf r}$ using a $%
V_{T_{s}}({\bf r;[}\rho ])$ derived from an approximate $T_{s}[\rho ]$ in
eq (\ref{kineticpotential}) can be poorly conditioned: very large errors in
the density at particular values of ${\bf r}$ can arise from an approximate $%
T_{s}[\rho ]$ whose numerical value differs by only a relatively small
amount from the correct value. For example, as discussed below, a
self-consistent solution of eq (\ref{eq:4}) using a $V_{T_{s}}({\bf r;[}
\rho ])$ arising from the local Thomas-Fermi approximation to $%
T_{s}[\rho ]$ and a $V_{eff}({\bf r})$ appropriate for an Ar atom predicts
an {\em infinite} density at the nucleus, although the numerical value for $%
T_{s}$ is only about 30\% above the correct value.
                                                                                                                    
But there is another way to make progress. If we can find a reasonably
accurate way to calculate $\rho ({\bf r})$ directly from eq (\ref{eq:4})
for a variety of $V_{ext}({\bf r}),$ then we can integrate the induced
density changes as the field is ``turned on'' to obtain the total energy $E$
and $T_{s}.$ See eq (\ref{eq:19}). Now the smoothing property of the
integration helps make the results for $E$ less sensitive to whatever errors
remain in $\rho ({\bf r})$ and we would expect good results \cite{Chen}.
                                                                                                                    
However, in general the integration must be carried out numerically over a
range of density changes and simple analytical approximations to $V_{T_{s}}(%
{\bf r;[}\rho ])$ do not automatically translate into equivalently simple
analytical expressions for $T_{s}[\rho ].$ If accurate direct approximations
for $T_{s}[\rho ]$ could be found the corresponding determination of $E$ and
$T_{s}$ would require only the (final) equilibrium density $\rho ({\bf r})$
and would certainly be simpler. It remains to be seen whether the benefits
of focusing first on $V_{T_{s}}({\bf r;[}\rho ])$ will outweigh this cost.
In the interim our approach still preserves many of the potential advantages
of OF methods over the more complicated KS procedure of determining $T_{s}$
numerically using orbitals, and we believe that insights gained from its use
may lead to better direct approximations for $T_{s}[\rho ].$
                                                                                                                    
Similar ideas has been successfully applied to classical nonuniform
fluids \cite{Chen, Weeks, Katsov, Katsov2},
where the density response and free energy change arising from a wide
variety of external fields $V_{ext}({\bf r})$ have been accurately
determined from the classical analogue of eq (\ref{eq:4}).
The terms in that equation can be given physically
suggestive interpretations and the results are comparable to those arising
from the best versions of classical weighted density functional theories. In
the simplest version of this theory, appropriate for very slowly varying
effective fields, one arrives at a (local) differential equation involving
only the density and a {\em linear} second derivative term, as in the
classical van der Waals theory of the liquid-vapor interface \cite{Weeks}.
Nonlocal
effects arising from more rapidly varying fields can be taken into account
directly in the density response by using an optimized version of linear
response theory \cite{Weeks, Chen, Katsov, Katsov2}.
                                                                                                                    
In this paper we consider only local approximations to $V_{T_{s}}({\bf r;[}
\rho ])$ that at each ${\bf r}$ involve the local density $\rho ({\bf r})$
and its lowest order gradients, motivated by analogous approximations made
for $T_{s}[\rho ]$ in the Thomas-Fermi theory and the von Weizs\"{a}cker
modifications and by the equivalent theory for classical fluids. In future
work we plan to examine nonlocal corrections suggested by an optimized
version of linear response theory and thus make contact with
current work on OF methods and earlier related work by Pratt and
coworkers \cite{Pratt}.
                                                                                                                    
\section{Thomas-Fermi and von Weizs\"{a}cker approximations for $T_{s}[\rho
] $}
                                                                                                                    
The first direct approximation for the kinetic energy functional was found
in the Thomas-Fermi (TF) model \cite{Thomas,Fermi}, where
\begin{equation}
T_{TF}[\rho ]=C_{F}\int \rho ^{\frac{5}{3}}({\bf r})d{\bf r.}  \label{eq:7}
\end{equation}
Here $C_{F}=\frac{3}{10}(3\pi ^{2})^{\frac{2}{3}}$. The TF kinetic energy
functional, $T_{TF}[\rho ]$, is derived by local use of a uniform free
electron gas model and is known to be exact in the limit of an infinite
number of electrons \cite{Lieb}.
                                                                                                                    
However, for an atomic system, the number of electrons $N$ is finite and the
effective potential is very rapidly varying near the nucleus. As a result,
there are many deficiencies in the TF model for such applications. The total
energies predicted for atomic systems give at best an order of magnitude
estimate, and cannot be relied on for quantitative calculations. More
serious errors can be seen in the density response at particular values of $%
{\bf r}$ as determined from the Euler equation (\ref{eq:4}). As mentioned
above, the TF model predicts an infinite electron density at the nucleus. In
addition, the density does not decay exponentially in the classically
forbidden region (the tail region), and it does not show any shell
structure. Furthermore, the TF model predicts no atomic bonding to form
molecules or solids \cite{Teller, Balazs}.
                                                                                                                    
Though the TF model has many shortcomings, it is a natural local approximation
and gives a simple analytical expression for the kinetic energy density
functional and the kinetic potential. Moreover, it is known that greatly
improved results can be obtained for atomic and molecular ground state
energies \cite{Lacks, Iyengar, Wesolowski, Thakkar}
if better quality densities are used in $T_{TF}[\rho ]$ instead of
the self-consistent variational densities arising from a solution of eq (%
\ref{eq:4}). For example, if the
Hartree-Fock (HF) density \cite{Clementi} for Ar is used
in $T_{TF}[\rho ]$, the kinetic
energy is only 7\% below the correct result. This suggests that improvements
to the TF model should focus directly on the Euler equation (\ref{eq:4}) and
the density response it predicts.
                                                                                                                    
In this paper we show that the Euler equation (\ref{eq:4}) can give
surprisingly
accurate results for atomic densities and energies if the kinetic potential
derived from the original TF model is augmented by a simple gradient term
that helps satisfy important boundary conditions both near and far away from
the nucleus. The specific form of this correction was inspired by the von
Weizs\"{a}cker modification of the TF kinetic energy density functional,
which we now describe.
                                                                                                                    
By considering modified plane waves of the form $(1+{\bf a}\cdot {\bf r})e^{i%
{\bf k}\cdot {\bf r}}$, with ${\bf a}$ a constant vector and ${\bf k}$ the
local wave vector, von Weizs\"{a}cker \cite{Weizsacker} derived a correction
\begin{equation}
T_{W}[\rho ]=\frac{1}{8}\int \frac{|\nabla \rho ({\bf r})|^{2}}{\rho ({\bf r}%
)}d{\bf r}  \label{Twcorrection}
\end{equation}
to the TF kinetic energy functional. A generalized total kinetic
energy functional
incorporating a correction of this kind can be written as
\begin{equation}
T_{TF\lambda W}[\rho ]=T_{TF}[\rho ]+\lambda T_{W}[\rho ],  \label{eq:8}
\end{equation}
where $\lambda $ is a parameter. $\lambda =1$ corresponds to the original
von Weizs\"{a}cker correction, and its self-consistent use in eq (\ref{eq:4})
for atoms not only yields a finite electron density at the nucleus with
the exact cusp condition \cite{Kato} satisfied, but also gives the
correct exponential
decay of the density far from the nucleus \cite{Chan, Handy} for the
exact $\mu $.
We refer to this as the TFW model. Disappointingly, however, the kinetic energy
given by this approach is not much better than that of the original
TF model. For
example, the results for Ar are 29\% below the correct value.
                                                                                                                    
Other values of $\lambda $ have also been examined. All TF$\lambda $W models
predict a finite electron density at the nucleus, and exponential decay far
from nucleus but only $\lambda =1$ satisfies the exact cusp condition and
gives the correct exponential decay constant. $\lambda =1/9$ has been shown
to give the correct second order gradient expansion correction to the TF
functional and to reproduce the exact linear response of a homogeneous
electron gas under a long wavelength perturbation \cite{Yang}, but energy
predictions for atoms again are inaccurate. Tomishima and Yonei \cite
{Tomishima} found empirically that rather accurate atomic ground state
energies for a wide range of $Z$ can be obtained by using $\lambda =1/5.$
However the inaccurate description of the electron density near and far from
nucleus and the absence of shell structure are still present, and part of
this good agreement probably comes from a cancellation of errors.
                                                                                                                    
These models use simple gradient corrections suggested by the von
Weizs\"{a}cker theory to try to improve the TF kinetic energy functional
directly. However nonlocal effects certainly must be important in
$T_{s}[\rho ]$, as is recognized in recent OF methods, and these must be
taken into account before this approach is likely to be quantitatively
useful.
                                                                                                                    
\section{Modified Euler equations for the density response}
                                                                                                                    
As argued above, it may be more profitable to examine simple gradient
corrections directly in the more local Euler equation. The Euler equation
arising from use of the TF$\lambda $W expression for $T_{s}[\rho ]$ is
\begin{equation}
\mu =\frac{5}{3}C_{F}\rho ^{\frac{2}{3}}({\bf r})+\frac{\lambda }{8}\frac{
|\nabla \rho ({\bf r})|^{2}}{\rho ({\bf r})^{2}}-\frac{\lambda }{4}\frac{
\nabla ^{2}\rho ({\bf r})}{\rho ({\bf r})}+V_{eff}({\bf r})  \label{eq:9}
\end{equation}
There is a local term in the density from the TF model and {\em two} terms
involving density gradients. Both terms must be present with specified
coefficients if we insist on using simple local expressions of the von
Weizs\"{a}cker form directly in $T_{s}[\rho ]$ with no compensating terms
from nonlocal corrections. However the first nonlinear gradient term does
not have a simple physical interpretation and does not appear in the
analogous classical van der Waals equation for the liquid-vapor interface
\cite{Weeks}. A linear gradient correction like the second term would
be expected
on quite general grounds. Moreover we find that the second term alone
is responsible
for producing a finite density at an atomic nucleus with $V_{ext}({\bf r}
)=V_{ext}(r)=-Z/r$, effectively building in the uncertainty principle,
and of course this term alone contributes to the linear
response of the density to weak fields. Finally this term yields exponential
decay far from the nucleus.
                                                                                                                    
Thus we propose as the simplest correction to the TF kinetic potential a
modified Euler equation where only the linear gradient term appears, with a
coefficient chosen to best describe the density response. To avoid confusion
in notation, $\lambda $ is changed to $\alpha $ in our new Euler equation:
\begin{equation}
\mu =\frac{5}{3}C_{F}\rho ^{\frac{2}{3}}({\bf r})-\frac{\alpha }{4}\frac{%
\nabla ^{2}\rho ({\bf r})}{\rho ({\bf r})}+V_{eff}({\bf r}).  \label{eq:12}
\end{equation}
Note that we use a $V_{eff}({\bf r})$ that includes the effects of exchange and
correlation in all the calculations reported herein. Thus Eq.\ (\ref{eq:12})
does not reduce to the classical TF model when $\alpha =0$.
                                                                                                                    
We first examine the predictions of this model for atoms. It is easy to show
that the density at the nucleus is finite for any choice of $\alpha $ and
satisfies
\begin{equation}
\frac{\rho ^{\prime }(0)}{\rho (0)}=-\frac{2Z}{\alpha }.  \label{eq:13}
\end{equation}
The exact cusp condition \cite{Kato} for the logarithmic derivative
of the density at
the origin can be satisfied if $\alpha =1$, as in the TFW model.
However this choice does not guarantee correct values for $\rho (0)$ and $\rho
^{\prime }(0)$ separately, and indeed the TFW model gives very poor results
for these terms. For example, for Ar $\rho (0)$ is 72\% below the HF value.
Moreover the nonlinear gradient term in the TFW model does not seem to
improve the density at intermediate values of $r$ when compared to the
results of our model.
                                                                                                                    
Equation (\ref{eq:12}) also produces exponential decay of the density at
large $r$, where
\begin{equation}
\rho (r)\approx \exp [-(-4\mu /\alpha )^{\frac{1}{2}}r].  \label{eq:14}
\end{equation}
When $\alpha =1/2$, the correct exponential decay constant for the exact
$\mu $ is predicted.
                                                                                                                    
Using this one parameter model we cannot satisfy both the exact cusp
condition and asymptotic decay condition with the same $\alpha $, and it is
clear that no oscillatory shell structure will be generated by the simple
gradient term. (As is the case for the classical van der Waals theory of
interfaces, nonlocal corrections are required to get oscillatory
structure \cite{Weeks}.)
But we can hope that a proper choice of $\alpha $ could give a
smoothed electron density that reproduces important overall features and
gives improved atomic energies when compared to the TF model. Since any
$\alpha $ produces a finite density at the nucleus and thus corrects a major
shortcoming of the TF model, we choose $\alpha =1/2$ to give the correct
exponential decay of the density far from the nucleus. This is the region
that should be most important in chemical bonding. In the following we refer
to eq (\ref{eq:12}) with $\alpha =1/2$ as the modified Thomas-Fermi (MTF)
model. We show below that the MTF model gives atomic energies comparable to
those of the best TF$\lambda $W model with $\lambda =1/5$ along with a much
better description of the density distribution.

\section{Results for the MTF model}

\subsection{Linear response function}

We first examine the predictions of the various models for the linear
response of the density of the model system to a weak perturbing potential.
In Fourier space this can be written as
\begin{equation}
\delta \rho({\bf k})=\chi({\bf k})\delta V({\bf k}).
\label{eq:16}
\end{equation}
Here $\chi({\bf k})$ is the Fourier transform of the linear response
function for a uniform system of noninteracting electrons, which can be
derived exactly from first-order perturbation theory \cite{Lindhard}:
\begin{equation}
\chi({\bf k})=-\frac{k_{f}}{\pi ^{2}}(\frac{1}{2}+\frac{1-q^{2}}{
4q }\ln |\frac{1+q }{1-q }|),  \label{eq:17}
\end{equation}
where $k_{f}$ is the Fermi wave vector and $q =k/2k_{f}$ is a
dimensionless momentum.
                                                                                                                    
The corresponding response function for the TF${\lambda }$W model, eq (\ref
{eq:8}), is given by \cite{Jones}
\begin{equation}
\chi({\bf k})_{TF\lambda W}=-\frac{k_{f}}{\pi ^{2}}(\frac{1}{%
1+3\lambda q ^{2}}),  \label{eq:18}
\end{equation}
as can be seen from eq (\ref{eq:9}) for an infinitesimal $V_{eff}.$ It is
easy to see that the nonlinear gradient term in (\ref{eq:9}) does not
contribute
to the linear response function. As a result, the linear response function
for the MTF model, derived from eq (\ref{eq:12}) with $\alpha =1/2,$ is
the same as the TF$\frac{1}{2}$W model.

\begin{figure}
\begin{center}
\includegraphics[width=80mm,height=60mm]{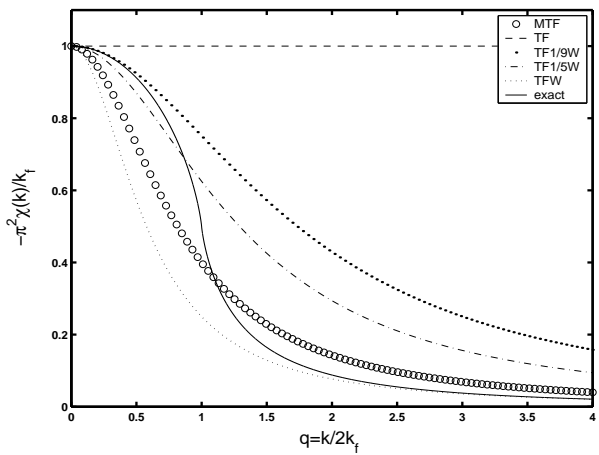}
\end{center}
\caption{\label{1} Linear response functions of a uniform system of independent
fermions as given by the various models.}
\end{figure}

As shown in Figure 1, the TF$\frac{1}{9}$W model is exact for long-wavelength
perturbations, but becomes worse in the high-$k$ region, while the TFW model
is exact for short-wavelength perturbations, but fails in the low-$k$
regions. The MTF model gives a qualitatively reasonable average description
of the exact linear response function, especially in the important region
near the singularity at $q =1$. Wang et al. \cite{YWang3} have suggested
that reproducing the singularity and the overall form of the linear response
function are important for obtaining the correct physics from OF methods,
particularly for producing the shell structure of atoms. While no simple
gradient type model can describe the singularity exactly (or give shell
structure), the MTF model captures the average behavior of the linear
response function and shows better accuracy in the physically significant
region near $q =1$.

\subsection{Atoms}

We carried out benchmark calculation on the hydrogen atom and the rare gas
atoms based on our MTF model. Our results are compared with the TF model,
and its von Weizs\"{a}cker-type gradient corrections. We examined $\lambda
=\frac{1}{9}, \frac{1}{5}, \frac{1}{2},$ and $1$ but report results only
for $\lambda =\frac{1}{5}$ and $1$, and for the KS and HF methods. The
local density approximation (LDA) \cite{Dirac, Ceperley, Ceperley2, Perdew}
for the exchange-correlation functional is used for all the models. The
standard method for solving Euler equations derived from the TF$\lambda $W
model \cite{Abrahams} was implemented to solve the Euler equation for the
MTF model and the KS theory. Our TF$\frac{1}{5}$W and KS results are
in very good
agreement with previous calculations \cite{Garcia-Gonzalez} using a
slightly different LDA
parametrization \cite{Perdew2}. The code uses the finite difference method with
the Gauss-Chebyshev (of the second kind) radial quadrature, proposed by Becke
{\em et al.}, consisting of 1000 points, and Becke's algorithms for solving
Poisson's equation for the Hartree potential \cite{Becke}.
                                                                                                                    
As shown in Table \ref{table:1}, the electron density at the nucleus, $\rho
(0),$ calculated by the MTF model is close to the HF results. As
discussed above, the TF model predicts an infinite
value for the electron density, while the TF$\frac{1}{5}
$W models overestimates the density by a factor of 4.
The TFW model considerably underestimates the density, despite satisfying
the exact cusp condition. It is known \cite{Cioslowski} that $\rho (0)$
is related to the density moments
$\langle r^{-1}\rangle $ and $\langle r^{-2}\rangle $.
Thus the MTF model would be expected to give good
results for these expectation values, as is shown in Table \ref{table:2}
and \ref{table:3} respectively.

\begin{table}
\caption{\label{table:1} Electron density at the nucleus $\rho(0)$ using the various TF-type models,
the KS method, the HF method and the MTF model.}
\begin{tabular} {c c c c c c}
\hline \hline
                                                                                                             
$ $&$MTF$&$HF\footnotemark[1]$&$TF\frac{1}{5}W$&$TFW$&$KS$ \\ \hline
$H$    & $0.5820$ & $0.3183$ & $2.390$ & $0.1029$ & $0.2728$ \\ \hline
$He$   & $5.208$  & $3.596$  & $19.29$ & $0.9515$ & $3.525$  \\ \hline
$Ne$   & $788.2$  & $619.9$  & $2595$  & $169.6$  & $614.5$  \\ \hline
$Ar$   & $4811$   & $3840$   & $15482$ & $1093$   & $3819$  \\ \hline
$Kr$   & $40064$    & $32236$   & $126491$  & $9579$  & $32146$  \\ \hline
$Xe$   & $137631$   & $112219$  & $431005$  & $33710$  & $111005$   \\ \hline
\hline
\end{tabular}
\footnotetext[1]{the HF data is taken from Clementi {\em et al} \cite{Clementi}.}
\end{table}

\begin{table}
\caption{\label{table:2}  $\langle r^{-1}\rangle$ using the TF$\frac{1}{5}$W model, the KS method,
and the MTF model.}
\begin{tabular} {c c c c}
\hline \hline
                                                                                                            
$ $&$MTF$&$TF\frac{1}{5}W$&$KS$\\ \hline
$H$&$1.000$&$1.249$&$0.9208$   \\ \hline
$He$&$2.907$&$3.261$&$3.312$ \\ \hline
$Ne$&$30.56$&$30.75$&$31.00$   \\ \hline
$Ar$&$69.93$&$69.63$&$69.62$   \\ \hline
$Kr$&$183.5$&$181.7$&$182.7$   \\ \hline
$Xe$&$321.2$&$317.8$&$317.8$   \\ \hline
\hline
                                                                                                            
\end{tabular}
\end{table}

\begin{table}
\caption{\label{table:3}  $\langle r^{-2}\rangle$ using the TF$\frac{1}{5}$W model, the KS method,
the HF method, and the MTF model.}
\begin{tabular} {c c c c c}
\hline \hline
                                                                                                             
$ $&$MTF$&$TF\frac{1}{5}W$&$KS$&$HF\footnotemark[1]$\\ \hline
$H$&$2.394$&$4.720$&$1.749$&$-$   \\ \hline
$He$&$11.50$&$20.06$&$11.71$&$11.99$ \\ \hline
$Ne$&$394.7$&$599.7$&$411.9$&$414.9$   \\ \hline
$Ar$&$1390$&$2056$&$1459$&$1465$   \\ \hline
$Kr$&$6025$&$8705$&$6318$&$-$   \\ \hline
$Xe$&$14101$&$20148$&$14799$&$14818$   \\ \hline
\hline
                                                                                                             
\end{tabular}
\footnotetext[1]{the HF data is taken from Porras {\em et al} \cite{Porras}.}
\end{table}

In Table \ref{table:4} we compare the chemical potential $\mu $ for the MTF
theory, calculated by a self-consistent solution of eq (\ref{eq:12}), with
that given by the KS equation and find very good agreement. This means that
the asymptotic behavior of the density predicted by the MTF model is close
to the prediction of the KS theory, since the MTF model would give
exact results
for the exact $\mu .$ Since the accuracy of the density in the tail is an
important indicator of chemical bonding \cite{Parr}, we believe that
this desired
behavior could lead to chemical bonding in molecules.

\begin{table}
\caption{\label{table:4} Atomic chemical potential $\mu$ using the various TF-type models, the KS 
method, and the MTF model.}
\begin{tabular} {c c c c c c}
\hline \hline
                                                                                                            
$ $&$MTF$&$TF$&$TF\frac{1}{5}W$&$TFW$&$KS$   \\ \hline
$H$   &$-0.2825$&$-0.08039$&$-0.09549$&$-0.09773$&$-0.2337$  \\ \hline
$He$  &$-0.3822$&$-0.06290$&$-0.1013$&$-0.1393$ &$-0.5702$  \\ \hline
$Ne$  &$-0.4400$&$-0.04959$&$-0.1093$&$-0.2188$ &$-0.4978$  \\ \hline
$Ar$  &$-0.4304$&$-0.05034$&$-0.1112$&$-0.2346$ &$-0.3823$  \\ \hline
$Kr$  &$-0.4224$ &$-0.05017$&$-0.1130$&$-0.2485$ &$-0.3464$  \\ \hline
$Xe$  &$-0.4192$&$-0.05034$&$-0.1138$&$-0.2552$ &$-0.3100$  \\ \hline
\hline
                                                                                                            
\end{tabular}                                                                                               
\end{table}

The Euler equation (\ref{eq:12}) for the MTF model is not derived by a
functional derivative of an energy functional. As a result we cannot
immediately obtain the energy once we have determined the density, as would
be possible in OF methods that give direct approximations for $T_{s}[\rho].$
Instead, we can use a coupling parameter approach \cite{Lee, Dominicis,
Rajagopal, Nalewajski, Pratt} to calculate the energy change as the external
potential is turned on:
\begin{equation}
E=E_{\lambda =0}+\mu \int [\rho _{\lambda =1}({\bf r})-\rho _{\lambda =0}(%
{\bf r})]d{\bf r}+\int_{0}^{1}d\lambda \int \rho _{\lambda }({\bf r})V_{ext}d%
{\bf r}  \label{eq:19}
\end{equation}
Here the external potential $V_{ext}$ is linearly scaled by a coupling
parameter $\lambda $ (not to be confused with the $\lambda $ parameter in
the TF$\lambda W$ methods) and the corresponding density
is $\rho _{\lambda }$. $E$
is the total energy of the system, and $E_{\lambda =0}$ is the total energy
at the same $\mu $ in the absence of the external potential.
                                                                                                                    
If exact values
for $\rho _{\lambda }$ were used, eq (\ref{eq:19})\ would give exact
results and a variety of different integration pathways (e.g., nonlinear
scaling of $V_{ext}$) could be used. Since we make approximations there will
be errors in our predicted $E$ values and different pathways could
(incorrectly) give different results.
However, our experience for the analogous classical problem suggests that these
errors are small if the density is reasonably accurate \cite{Chen},
as seems to be the case
here.
                                                                                                                    
We note that $\rho _{\lambda =0}$ and thus $E_{\lambda =0}$ vanish for the
uniform electron gas
because of the fixed negative value of the chemical potential.
Therefore, $E$, the total energy, can be straightforwardly computed
from eq (\ref{eq:19}) by numerical integration over a series of $\lambda $
values. For the benchmark calculations reported here for the total energy,
1000 points are used
between $\lambda =0$ and $\lambda =1$, though reasonable results can
be obtained with
about 10 points. The kinetic energy can then be found by subtracting
the potential
energy (calculated from the potential energy density functionals)
from the total energy.
This way of obtaining kinetic energy is equivalent to integrating the
kinetic potential directly as discussed above, but we believe it is numerically
more accurate. As shown in Table \ref{table:5}, the total energy of the MTF
model is comparable to that of the TF$\frac{1}{5}$W model, the best TF$%
\lambda $W model for the total atomic energy.

\begin{table}
\caption{\label{table:5} Atomic energy E using the various TF-type models, the KS method,
and the MTF model.}
\begin{tabular} {c c c c c c}
\hline \hline
                                                                                                            
$ $&$MTF$&$TF$&$TF\frac{1}{5}W$&$TFW$&$KS$ \\ \hline
$H$ &$-0.6092$&$-1.076$&$-0.6085$&$-0.2927$&$-0.4459$   \\ \hline
$He$&$-2.902$&$-4.756$&$-2.917$&$-1.559$&$-2.834$   \\ \hline
$Ne$&$-129.5$&$-176.6$&$-129.5$&$-86.40$&$-128.2$   \\ \hline
$Ar$&$-529.0$&$-681.1$&$-526.2$&$-375.5$&$-525.9$   \\ \hline
$Kr$&$-2774$&$-3378$&$-2747$&$-2099$&$-2750$   \\ \hline
$Xe$&$-7293$&$-8643$&$-7214$&$-5701$&$-7229$   \\ \hline
\hline
                                                                                                            
\end{tabular}
\end{table}

In Figure 2-5, we compare the radial density distribution $r^{2}\rho
(r) $ of the MTF model to that predicted by other theories. Although the MTF
model shows no shell structure when compared to the essentially exact
KS theory,
it gives a nice averaging
of the electron density, with slightly more emphasis on the principle peak
than in the TF$\lambda $W models, along with significantly better results both
near and far from the nucleus. The fact that a smooth density can give
such accurate energies illustrates the point that the integration in eq (%
\ref{eq:19}) renders it less sensitive to small errors in the density.

\begin{figure}
\begin{center}
\includegraphics[width=80mm,height=60mm]{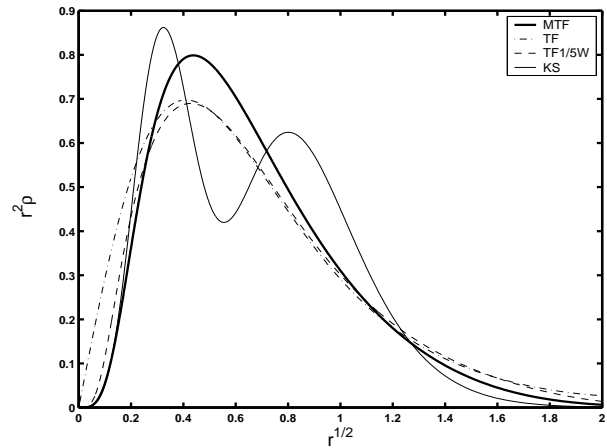}
\end{center}
\caption{\label{2}  Radial density $r^2 \rho$ of the Neon atom with the various TF-type
models, the KS method, and the MTF model.}
\end{figure}

\begin{figure}
\begin{center}
\includegraphics[width=80mm,height=60mm]{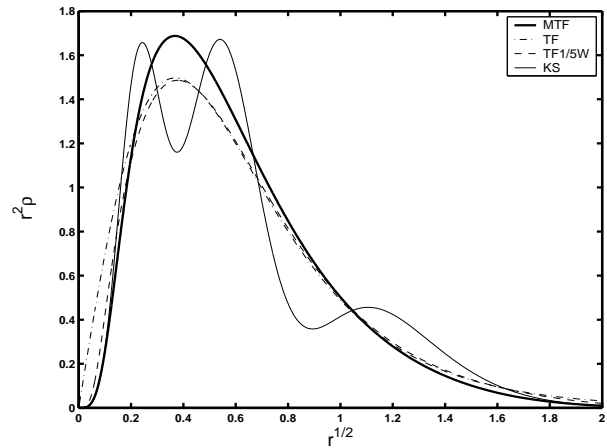}
\end{center}
\caption{\label{3} Same as in Figure 2 but for the Argon atom}
\end{figure}

\begin{figure}
\begin{center}
\includegraphics[width=80mm,height=60mm]{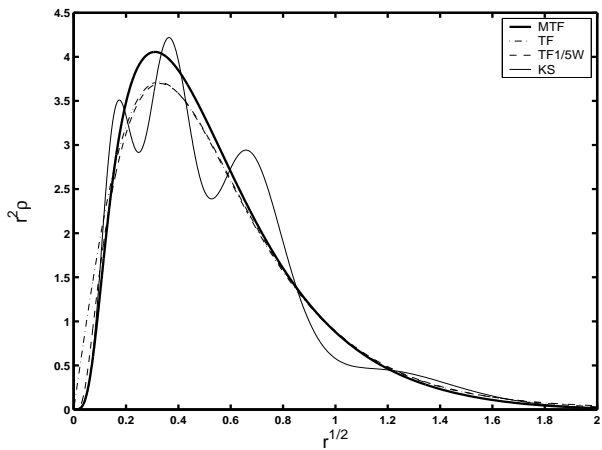}
\end{center}
\caption{\label{4} Same as in Figure 2 but for the Krypton atom}
\end{figure}
                                                                                                             
\begin{figure}
\begin{center}
\includegraphics[width=80mm,height=60mm]{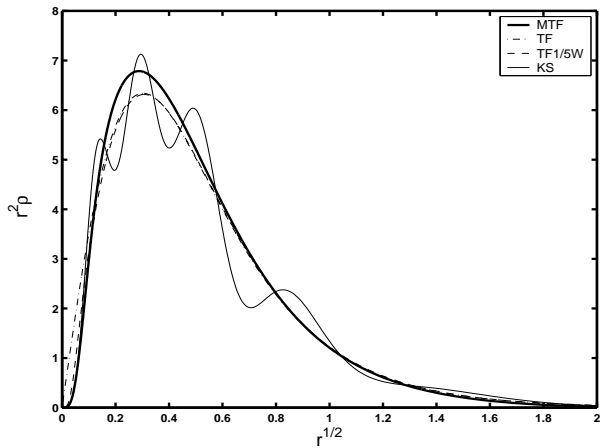}
\end{center}
\caption{\label{5} Same as in Figure 2 but for the Xenon atom}
\end{figure}

\subsection{Diatomic molecules}

Gordon and Kim (GK) suggested a simple model for calculating the
intermolecular potential of closed-shell molecules \cite{Gordon}. By
assuming the molecular density of a diatomic molecule can be expressed as
the sum of the two separate atomic densities, reasonable bonding potentials
for closed-shell molecules can be obtained. Instead of using the inaccurate
TF atomic densities, GK used the atomic HF densities and inserted
these into the
original TF kinetic energy functional, together
with the classical electron-electron and nuclear-electron potentials,
and a local density approximation for the exchange-correlation functionals.
                                                                                                                    
We followed GK's simple approach, using atomic densities for the Ar
atom given by the various models discussed here. Becke's
atomic partition, the Gauss-Chebyshev (of the second kind) radial quadrature,
and the Clenshaw-Curtis angular quadrature with 1000 radial points and 500
angular points are used in the numerical calculations \cite{Becke2,
Perez-Jorda}.

\begin{figure}
\begin{center}
\includegraphics[width=80mm,height=60mm]{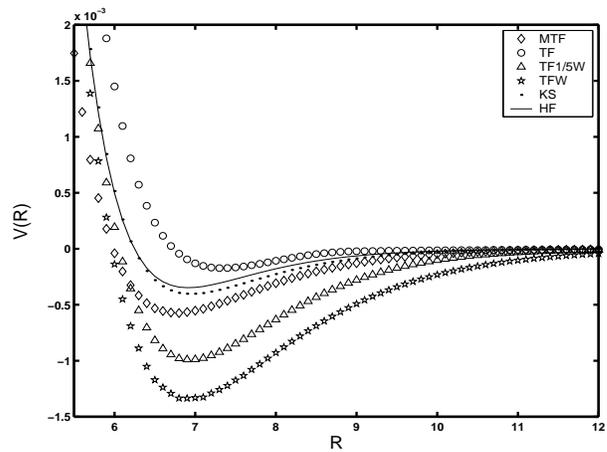}
\end{center}
\caption{\label{6} Ar-Ar interaction potential via the Gordon-Kim approach.}
\end{figure}

We found that the KS bonding potential is very similar to the HF result, as
is expected due to the high quality of its atomic density. As we can see in
Figure 6, the strength of the bonding of the TF$\lambda $W models increases
as $\lambda $ increases. However, the bonding energy is consistently
overestimated, even for the smallest value of $\lambda $. The MTF model predicts a
reasonable bonding curve, with both its strength and bond length close to
the HF and KS methods. Note that even the TF density can give a
qualitatively accurate bonding curve using the GK method, though a
self-consistent solution of the molecular density using the TF model is
known to give no bonding \cite{Teller}.

\section{Possible generalizations}

We used heuristic arguments to suggest the form of the MTF equation
(\ref{eq:12}).
One might hope that a more systematic search using generalized forms of this
kind could yield even better results. Since the MTF model has only one
parameter, $\alpha $, we considered one such generalization of the gradient
term in (\ref{eq:12}) that allows for the introduction of a second
parameter, $\beta $, and that interpolates between the MTF and the
TF$\lambda $W models:
\begin{equation}
-\frac{\alpha }{4}\frac{\nabla ^{2}\rho ^{\beta }}{\rho ^{\beta }}=-\frac{%
\alpha }{4}\left\{ \beta (\beta -1)\frac{|\nabla \rho ({\bf r})|^{2}}{\rho (%
{\bf r})^{2}}+\beta \frac{\nabla ^{2}\rho ({\bf r})}{\rho ({\bf r})}\right\}
.  \label{generalized gradient}
\end{equation}
For $\alpha =1/2$ and $\beta =1$, this reduces to the MTF form, while the TF$%
\lambda $W models correspond to $\{\alpha ,\beta \}=\{2\lambda ,1/2\}$, as
can be seen from eq (\ref{eq:9}).
                                                                                                                    
Using this generalized form in eq (\ref{eq:12}) for atomic systems, the
density at the nucleus is finite and satisfies a modified cusp condition
\begin{equation}
\frac{\rho ^{\prime }(0)}{\rho (0)}=-\frac{2Z}{\alpha \beta }.  \label{eq:25}
\end{equation}
When $\alpha \beta =1$, the exact cusp condition is satisfied. This form
also produces an exponentially decaying density in the tail, satisfying
\begin{equation}
\rho (r)\approx \exp \left[ -\left( \frac{-8\mu }{%
2\alpha \beta ^{2}}\right) ^{\frac{1}{2}}r\right] .  \label{eq:26}
\end{equation}
When $2\alpha \beta ^{2}=1$ the correct asymptotic decay constant is found
for the exact $\mu $.
                                                                                                                    
However only when $\{\alpha ,\beta \}=\{2,1/2\}$ (i.e., the original von
Weizs\"{a}cker model with $\lambda =1$) can both conditions be satisfied
simultaneously. Since all values give a finite density at the origin, we
also considered the one parameter family of models with $2\alpha \beta
^{2}=1 $ that give the correct exponential decay constant.
However, the best overall results for the
atomic energy and density were still found using the MTF model
with $\{\alpha ,\beta\}=\{1/2,1\}$.
Evidently the nonlinear gradient term in eq (\ref
{generalized gradient}) that is generated when $\beta \neq 1$ does not
improve the simple MTF model and other forms should be examined.
                                                                                                                    
\section{Final remarks}
                                                                                                                    
In summary, we have argued that it may be profitable to focus first on the
Euler equation determining the electron density when correcting the TF model
and more generally in developing OF approximations in DFT. Nonlocal effects
may be less severe, as is suggested by the generally good results found from
use of the simple gradient correction discussed here. In future work we hope
to take account of nonlocal corrections to the density response by using an
optimized version of linear response theory \cite{Weeks, Pratt}.
                                                                                                                    
As presently implemented in eq (\ref{eq:19}), a coupling parameter
integration is required to determine the total energy directly from the
density response. This is numerically more costly than would be the case if
accurate, variationally stable, direct OF approximations for $T_{s}[\rho ]$
can be found that yield good results for $\rho ({\bf r)}$ for all ${\bf r}$
when the associated $V_{T_{s}}({\bf r;[}\rho ])$ is used in the Euler
equation.
                                                                                                                    
A less ambitious goal that may prove almost as useful is to
develop approximate energy functionals that, while not variationally
accurate enough to use directly in the Euler equation, still give good
numerical results for the kinetic and total energies of particular classes
of systems if reasonably accurate densities are supplied as input. The use
of such functionals in theories like ours would eliminate the need for the
coupling parameter integration and simplify the calculation of the total
energy. As the simplest such example, if the MTF density is directly inserted
into the TF kinetic energy
functional, the deviation of the kinetic energy from the MTF value is less
than 10\% (except for the lightest atoms H (20\%) and He (17\%)). We found
empirically that the
TF$\frac{1}{7}$W functional gives results to within about 1\% for the heavier
atoms, though there is
no real justification for such a choice and a more systematic procedure is
called for. We are currently
investigating whether these ideas can be further developed and usefully applied
to molecules and extended systems.

\section{Acknowledgments}

This work has been supported by NSF grants CHE01-11104 (JDW, JDC) and
DMR01-04987 (JDC), and a NASA grant NCC8-152 (JDC).
Jeng-Da Chai acknowledges full support from the UMCP Graduate School Fellowship,
the IPST Alexander Family Fellowship, and the CHPH Block Grant Supplemental Fellowship.
We thank Yng-Gwei Chen for helpful discussions.

\bibliography{basename of .bib file}

\end{document}